# Cloud Based Application Development for Accessing Restaurant Information on Mobile Device using LBS

Keerthi S. Shetty and Sanjay Singh

Department of Information and Communication Technology Manipal
Institute of Technology, Manipal University, Manipal-576104, India
keert.cs@gmail.com, sanjay.singh@manipal.edu

**Abstract**

Over the past couple of years, the extent of the services provided on the mobile devices has increased rapidly. A special class of service among them is the Location Based Service(LBS) which depends on the geographical position of the user to provide services to the end users. However, a mobile device is still resource constrained, and some applications usually demand more resources than a mobile device can a ord. To alleviate this, a mobile device should get resources from an external source. One of such sources is cloud computing platforms. We can predict that the mobile area will take on a boom with the advent of this new concept. The aim of this paper is to exchange messages between user and location service provider in mobile device accessing the cloud by minimizing cost, data storage and processing power. Our main goal is to provide dynamic location-based service and increase the information retrieve accuracy especially on the limited mobile screen by accessing cloud application. In this paper w e present location based restaurant information retrieval system and w e have developed our application in Android.

## 1 Introduction

Mobile phones are becoming pervasive. Given the advances in mobile phones, users start to consider a mobile phone a personal information processing tool. That is, a user expects to execute any application on top of a mobile device. The information retrieval in mobile devices is a tedious task due to the limited processing capability and low storage space available. Therefore ways to explore technology where o oading to mobile devices can be overcome is a research issue. Hence the advent of Cloud computing in Location-Based Services increases the user's information retrieve capability by overcoming the mobile's storage space and processing capability. A lot of development in the eld of mobile computing devices can be seen during the recent years. With the rapid improvement in technology of these mobile computing devices, Cloud Computing has been gaining a lot of attention over the years. Cloud computing, a rapidly developing information technology, has aroused the concern of the whole world.

Cloud computing is Internet-based computing, whereby shared resources, software and information, are provided to computers and devices on-demand. It is not a new concept; it is originated from the earlier large-scale distributed computing technology [1]. However, it will be a subversion technology and cloud computing will be the third revolution in the IT industry, which represent the development trend of the IT industry from hardware to software, software to services, distributed service to centralized service.

Cloud computing is also a new mode of business computing, it will be widely used in the near future. The core concept of cloud computing is reducing the processing burden on the users' terminal by constantly improving the handling ability of the "cloud", eventually simplifying the users' terminal to a simple input and output devices, and busk in the powerful computing capacity of the cloud on-demand [1]. But any form of work in the eld of mobile devices accessing cloud service provider in LBS has been minimal.

Integration betw een mobile devices and cloud computing is presented in several previous w orks. Chris-tensen [2] presents general requirements and key technologies to achieve the vision of mobile cloud computing.

The author introduces an analysis on smart phones, context awareness, cloud and restful based web services, and explains how these components can interact to create a better experience for mobile phone users.

Luo [3] introduced the idea of using cloud computing to enhance the capabilities of mobile devices. The main goal of this work is to show the feasibility of such implementation, introducing a new partition scheme for tasks. The best point about this paper is the considerations about using the cloud to back mobile computing.

Giurgiu et al. [4] has used the cloud as the container for mobile applications. Applications are pre-processed based on the current context of the user, so only the bundles that can run on the local device and minimize the communication overhead with the cloud are o oaded to the mobile device from the cloud. They focus on partition policies to support the execution of application on mobile devices, and do not tackle any other issue related to mobile cloud computing.

Chun and Maniatis [5] have explored the use of cloud computing to execute mobile applications on behalf of the device. They propose the creation of clone VMs to run applications/services the same way that they will run on mobile devices in order to avoid inconsistencies produced to run part of a program in di erent architecture. Their work is strongly tied to distributed le systems, and assumes connectivity to the cloud.

In this paper, We present the application which has been implemented in Java for Android devices which require the Android SDK and ADT Plug-in. It was selected because it provides rich APIs for map, location functions and also there were implementations available for cloud computing providers on top of this platform. So we introduce the android operating system into our mobile information retrieve system. It can e ectively interact with cloud service providers to retrieve information in Location Based Services.

The remaining paper is organized as follows. Section 2 discusses the theoretical background of LBS and the message exchange that occurs in the system and cloud computing, android operating system. Section 3 describes the architecture of the system developed. Section 4 discusses about the system functionality of the system. Section 5 gives the algorithmic description. Section 6 describes the implementation details of the system. Finally, a conclusion has been drawn in section 7.

## 2 Theoretical Background

### 2.1 Location Based Services

Location Based Service(LBS) uses the geographical position of a user to provide services such as health, work, entertainment services etc. The mobile service provider are the entities that provide these services to the user.

A distinct characteristic of LBS is its capability to provide service not just based on time and location, but also based on the user requirement at a particular location. The LBS system should be aware of the user needs and capable of mapping it to the location at which the service is required. The complexity of this system increases when the accuracy of the position and the dependency relationships between the locations need to be considered [6].

There are various devices and techniques that can be used to detect the location of the user in the system. Some of the examples are Global positioning system (GPS), RFID etc.

#### 2.1.1 GPS based systems

The Global Positioning System is a navigation system that consists of 28 high-altitude satellites with highly accurate atomic clocks. These satellites are used to nd the precise geographical position of a user. The GPS services are usually freely available [7].

The GPS receiver uses a triangulation method of the satellites to pinpoint the location of a user. It can be used to nd the exact location to an order of a few meters. Error larger than a few meters is intolerable in these systems. GPS systems have a response time of the order of a few milliseconds making it an highly e cient system for LBS.

### 2.1.2 RFID systems

It is one of the technologies that has gained a lot of importance in the recent times. The distinct characteristics that separate it from other context aware technologies are contact-less, multi-object recognition, non-line-of-sight, long distance, large store of memory, programmability and penetrability [7]. The main advantage of RFID is its ability to map a physical object to a virtual object in its RFID network. This is achieved by assigning a physical tag to each physical object.

The entire area under the RFID system is divided into zones. These zones are then mapped into space of information tags. This mapping makes it easier to determine the accurate locations of the physical objects.

Currently, there is no system that is capable of giving the exact location information. GPS works accurately only on outdoors. It fails to provide satisfactory results when there is some kind of obstruction. Whereas on the other hand, RFID tags can be used on a request/response model to store unique RFID tags or some other form of identi er in their memory, and hence can be used to track mobile objects irrespective of their location.

## 2.2 Message exchange in LBS

When a user enters into the coverage area of a Location Service Provider(LSP), various messages are ex-changed. It could be the LSP sending a list of services to the client, or the user selecting among the list of services, or the messages could also include the LSP performing the authentication and authorization based on the information received from the mobile device. These messages exchanged form the backbone of the LBS system.

## 2.3 Cloud Computing

Cloud is a virtualized pool of computing resources. It can:

> Manage a variety of di erent workloads, including the batch of back-end operations and user-oriented interactive applications.

> Rapidly deploy and increase workload by speedy providing physical machines or virtual machines.

> Support for redundancy, self-healing and highly scalable programming model, so that workload can be recover from a variety of inevitable hardware/software failure.

> Real-time monitor resources usage, rebalance the allocation of resources when needed [8].

## 2.4 Android Operating System

The Open Handset Alliance released the Google Android SDK on November 12, 2007 [9]. The conception of the Android platform is attracting more and more programmers in mobile computing elds. Android is a package of software for mobile devices, including an operating system, middleware and core applications. The Android SDK provides powerful tools and APIs necessary to develop applications on the Android platform using the Java programming language. Android platform is of open system architecture, with versatile development and debugging environment, but also supports a variety of scalable user experience, which has optimized graphics systems, rich media support and a very powerful browser. It enables reuse and replacement of components and an e cient database support and support various wireless communication means. It uses a Dalvik virtual machine heavily optimized for mobile devices [10]. Android also supports GPS, Video Camera, compass, and 3D-accelerometer and provides rich APIs for map and location functions. Users can exibly access, control and process the free Google map and implement location based mobile service in his mobile systems at low cost. Android platform will not only promote the technology (including the platform itself) of innovation, but also help to reduce development costs, and enable developers to form their mobile systems with unique characteristics.

# 3 Architecture Description

The Fig.1 gives the architecture diagram of the cloud application. The application has been implemented in Java for android devices which require the Android SDK and ADT Plug-in. In this paper, we rst proposed a location-based data and service middleware, which is mainly responsible for the collection and disposal of di erent data type and services existing in di erent network information platform. Based on the pretreated information, this interface module will repackage the heterogeneous data and service and republic them as web service. The details of the cloud application layer are given in [11].

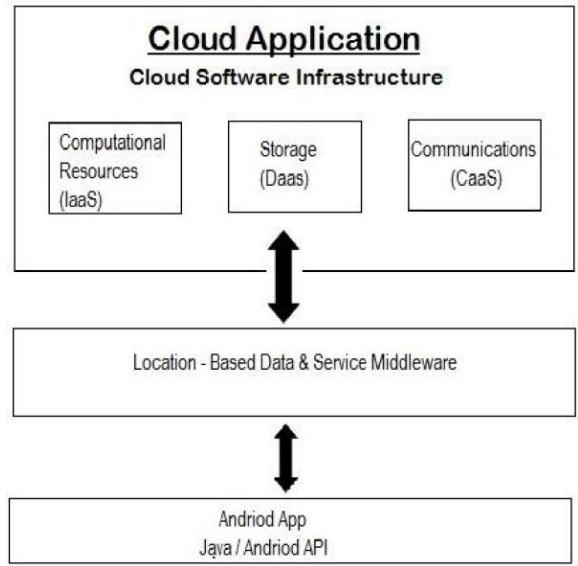

Figure 1: Architecture diagram of the cloud application

# 4 System Functionality

Each location has several Cloud Units(CU) which acts as mobile support station to support services for mobile users in this location. Cloud units in every location are connected to Cloud Service Provider(CSP). When a user arrives at a new city, he may want to know the typical or popular food or restaurant in this city. Then it is very hard for him to search. In our system, we have considered only restaurant searching services. Each cloud stores restaurant related information like address, contact number, food style information etc. The cloud enabled mobile application is shown in Fig.2.

## 4.1 Role of LSP and User

When a user enters into the coverage area of LSP, user needs to register with LSP to access the available services. LSP performs authentication by assigning user with unique ID i.e.,Phone Number. User is able to access required service by providing unique ID. Use case diagram of LSP and User is shown in Fig.3 and Fig.4 respectively.

## 4.2 Role of CSP and Cloud Units in LBS

In location based service, each location has several cloud units. Cloud units in every location are connected to CSP. These two have di erent computing ability and its main task is di erent. The cloud units aim at dealing the requests from users directly. The CSP is for dealing the key computing of some service. Cloud units in each location send requests to the CSP for some complex service. Use case diagram of CSP and CU is shown in Fig.5 and Fig.6 respectively.

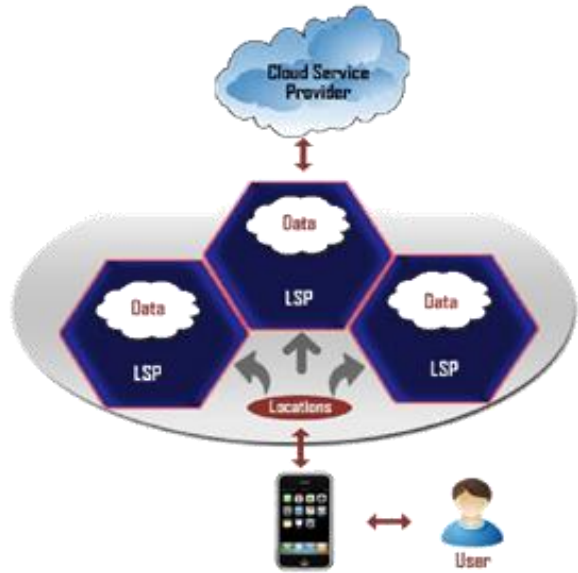

Figure 2: Cloud enabled mobile application

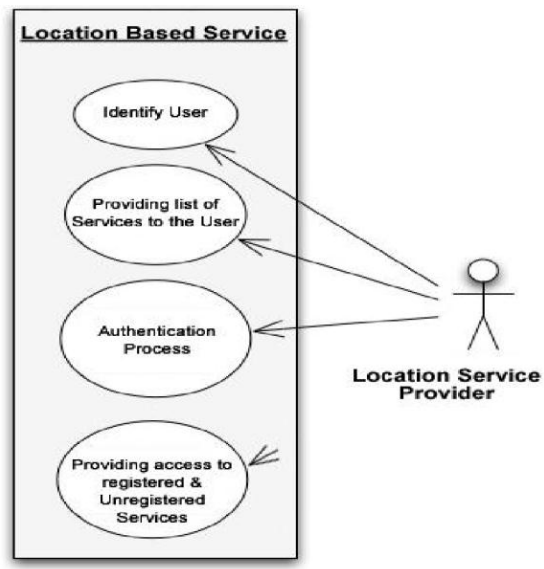

Figure 3: Use case diagram of Location Service Provider

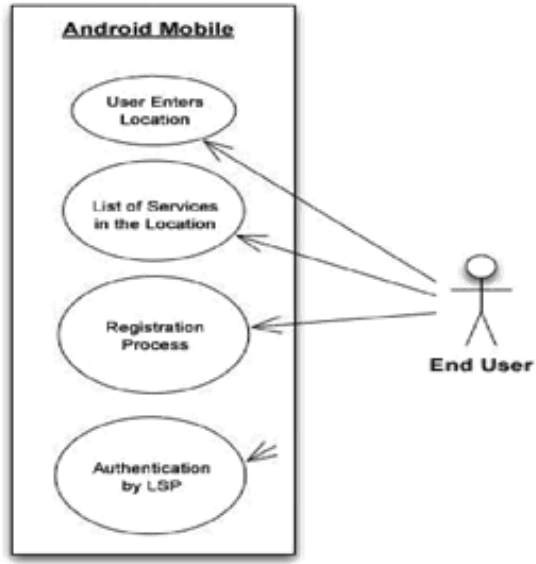

Figure 4: Use case diagram of User

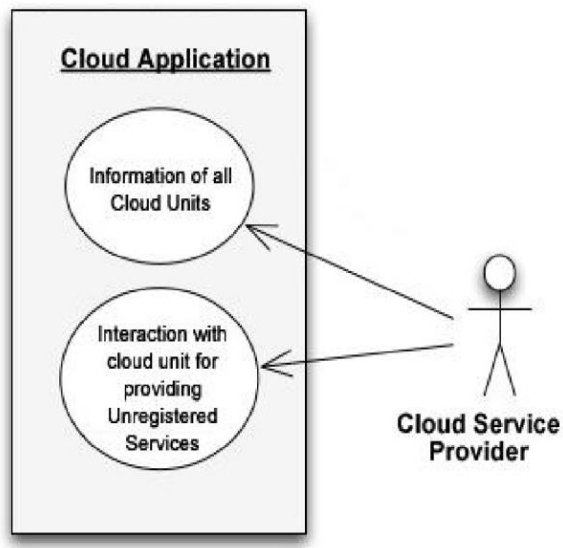

Figure 5: Use case diagram of Cloud Service Provider

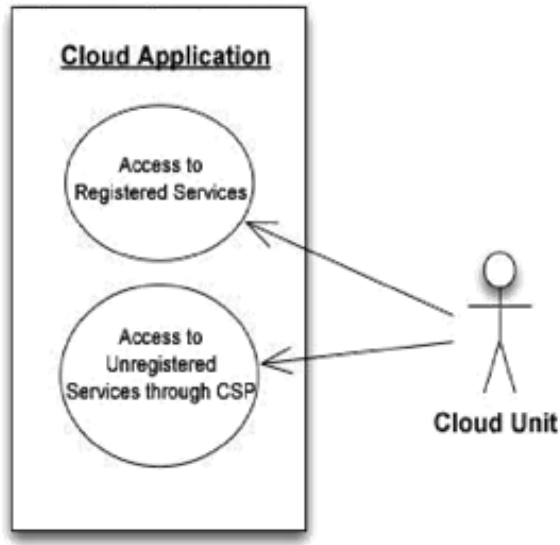

Figure 6: Use case diagram of Cloud Unit

# 5 Algorithmic Description

The LSP performs authentication when a user registers for its services. If the user is authenticated, the LSP provides the list of services to the user. The user selects a service among the services in this list. This list contains all the services the user is registered to and available in that location. Each location has cloud units capable of providing services to the user. User can access registered service from cloud units directly. User needs to contact cloud units present in each location for unregistered services. CU contacts CSP to provide required services to the user. This can be explained better with the help of a pseudo code given in Algorithm 1. The Algorithm 1 depicts the overall working of the system.

In Algorithm 2, the location of the user is identified using either RFID or GPS. Once the location has been identified, the number of users has to be computed. Once this computation is done, the location co-ordinates of the particular user is sent to the LSP.

In Algorithm 3, the location co-ordinates returned by Algorithm 2 is used as a parameter to identify all the restaurants available in that location.

In Algorithm 4, for the given list of restaurant services available in that location, the LSP checks whether the user is authorized to use the service or not. For all the services that the user is authorized, a list is created and sent to the user.

# 6 Implementation Details

Our main goal is to provide dynamic location-based service and increase the information retrieve accuracy especially in the limited mobile screen by accessing cloud application. The location is capable of providing multiple services. Each location has multiple cloud units. Cloud units stores information related to restau-rants like address, contact information and food style information etc. We have modularized the system into different modules of the system.

**Creating the CSP:** It stores information about all the cloud units present in each location.

**Registration Process:** When a user enters into the coverage area of service provider, user registers with service provider to access the available services based on his preferences. User enters his credentials

**Algorithm 1** LocationBasedRestaurantInfoRetrieveSystem()

1: RegisterUser (phoneno)
2: **for** every Registered user **do**
3:   when user enters any of the location
4:   Identify_Location()
5:   Identify_List_of_Restaurants_in_Location (location)
6:   **if** user is authenticated **then**
7:     **for** every user **do**
8:       Provide_RestaurantInfo (info)
9:     **end for**
10:     **for** every registered restaurant service selected by user **do**
11:       user access cloud unit present in location for restaurant info
12:       **for** every unregistered restaurant service selected by user **do**
13:         Cloud Unit Present in location contacts CSP
14:         CSP provides the service
15:       **end for**
16:     **end for**
17:   **end if**
18: **end for**

**Algorithm 2** Identify Location()

1: Identify the location the user is currently present using either RFID or GPS.
2: Identify the number of users present in the location for whom the services have to be provided.
3: **return** Location co-ordinates.

**Algorithm 3** Identify List of Restaurants in Location (location)

1: Identify the different restaurants that are available in the particular location.
2: Identify the what type of services to be provided to a specific user.
3: **return** Location co-ordinates.

**Algorithm 4** Provide RestaurantInfo (info)

1: **if** user is authorized for the restaurant service **then**
2:   add it to the list of available services.
3: **end if**
4: **return** List of Services

to get registered and the location service provider assigns him with the userID.

**Authentication performed by the speci c LSP:** The service provider performs authentication when a user registers for its services. If the user is authenticated, the service provider provides the access to the registered services. This list contains all the services the user is registered to in that location. The service provider provides the requested service to the user as long as the user is in that location.

**User access CU for the registered services:** When a user enters into the coverage area of service provider and gets registered with the LSP for his preferable services and gain access to the registered services. Each location has multiple cloud units which provide information to the registered user based on his request.

**User accessing CSP through the CU for unregistered services:** Registered user can directly interact with cloud units but not with cloud service provider. Cloud units in each location send requests to the cloud service provider for some complex service.

The application is developed for retrieving information about restaurant based on the current location. User can quickly nd the interesting restaurant. User can obtain the restaurant related information in that location.

The application will rst obtain the user's current location and show the name of the current location as well as list of restaurants under that location. User is provided with two options. Register button for new user to register with LSP for required information about restaurant. Login button for already registered user. It is shown in Fig.7.

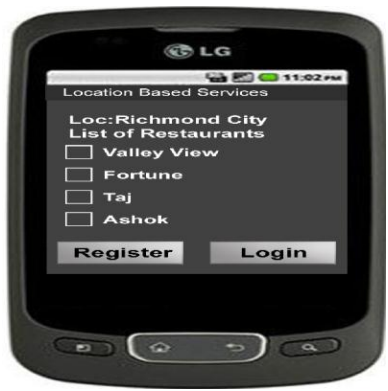

Figure 7: Screenshot showing list of restaurants under Location

When a user enters into the coverage area of LSP, user needs to register with LSP to access the available restaurant information in that location. User enters his credentials to get registered and LSP assigns user with unique ID. It is shown in Fig.8.

Registered user enters his login credentials to get access to the selected restaurant information. LSP performs authentication by unique ID assigned to each user, pop-up message is displayed "Authenticated User". It is shown in Fig.9.

Once the registered user login to the application the selected restaurant information is displayed. It is shown in Fig.10.

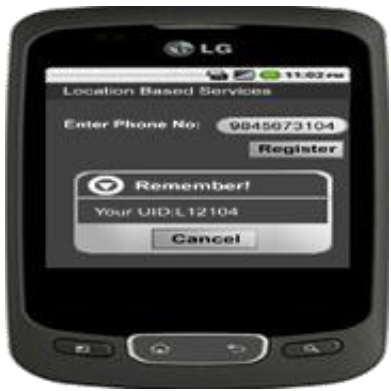

Figure 8: Screenshot showing registration process and UniqueId generation

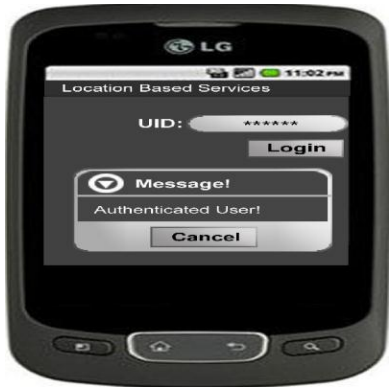

Figure 9: Screenshot showing login credentials and authentication process

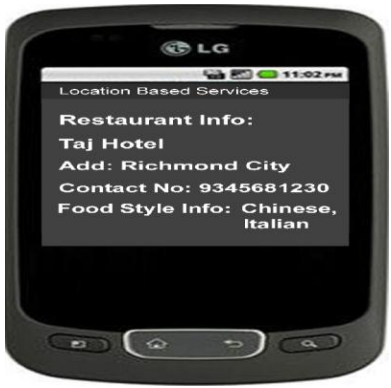

Figure 10: Screenshot showing selected restaurant information

# 7  Conclusion and Future Work

Providing dynamic location-based service and increase the information retrieve accuracy especially in the limited mobile screen have become the important research areas in the development of location-based services. Cloud computing brings us the approximately in nite computing capability, good scalability, service on-demand and so on. Cloud provides secure and dependable data storage center. In this paper, we have proposed and developed an Android based application to retrieve restaurant information in mobile device accessing CSP based on the locations. Registered user can access services directly from the CU whenever he wants irrespective of the location. Processing power is faster and it is energy e cient. The application was implemented in Java for Android devices. In the future, we believe that our e orts need to focus more on security issues and platform independent cloud application.